\documentclass{cimento}
\usepackage{graphicx}

\title{Equilibration dynamics and isospin effects in nuclear reactions}

\author{A.S.~Umar\from{ins:x}\thanks{email:umar@compsci.cas.vanderbilt.edu}, C.~Simenel\from{ins:y} \atque K.~Godbey\from{ins:x} }
\instlist{\inst{ins:x} Department of Physics and Astronomy, Vanderbilt University, Nashville, Tennessee 37235, USA
  \inst{ins:y} Department of Nuclear Physics, Research School of Physics and Engineering, The Australian National University,
        Canberra ACT 2601, Australia}
\shortauthor{A.S. Umar \etal}
\begin{document}

\maketitle

\begin{abstract}
We discuss equilibration times and isospin effect for various quantities in low-energy heavy-ion
reactions. These include equilibration of mass, isospin, and total kinetic energy (TKE) in quasifission and deep-inelastic
reactions. The
calculations are performed using the time-dependent Hartree-Fock theory.
The influence of shell effects on the equilibration times are also discussed
in the context of theoretical and experimental results.
\end{abstract}

\section{Introduction}
Low-energy heavy-ion reactions provide us a rich laboratory to study the equilibration
dynamics of strongly interacting many-body systems.
In addition, these reactions probe an intriguing interplay between the microscopic
single-particle dynamics and collective motion at time scales too short for complete
equilibration. In order to elucidate trends and systematics in these reactions
both theoretical and experimental studies must be undertaken for an assortment of
projectile and target combinations. Such studies is expected to be
under taken at the  current and future radioactive ion-beam (RIB) facilities~\cite{balantekin2014}.

In this manuscript we will discuss the equilibration dynamics and time-scales for various quantities that are
connected to the experimentally observable entities. These include the study of mass, isospin, and total
kinetic energy (TKE) equilibration time-scales. In most of these studies one is essentially dealing with the transport phenomena of
isospin asymmetric systems. Recently, charge equilibration, driven by the nuclear symmetry-energy,
has been experimentally studied near the Fermi energy~\cite{jedele2017}. At these energies the sticking
or contact times of the participating nuclei are sufficiently short to induce a partial charge equilibration~\cite{tsang2004}.
Charge equilibration has also been studied with deep inelastic collisions at lower energies, but with large isospin asymmetry
in the entrance channel~\cite{planeta1988,desouza1988b,planeta1990}.
In recent years a number of transport models have been employed to investigate the density
dependence of the symmetry energy away from the saturation
density~\cite{danielewicz2002,rizzo2008a,colonna1998,toro2010,rizzo2011,zhang2012,colonna2013}.
While considerable success has been achieved in obtaining information about the EOS from these
calculations more refinement of the models are needed to make a deeper connection to fundamental aspects
of nuclear many-body physics.

For the low-energy heavy-ion collisions the relative motion of the centers of the
two nuclei is characterized by a short wavelength and thus allows for a classical treatment, whereas
the wavelength for the particle motion is not small compared to nuclear sizes and should be treated
quantum mechanically. The mean-field approach such as the time-dependent Hartree-Fock (TDHF)
theory~\cite{simenel2012,simenel2018} and its extensions provide a microscopic basis for describing the
heavy-ion reaction mechanism at low bombarding energies. In this manuscript we provide studies of a
variety of nuclear reactions to address some of the issues discussed above.

\section{Equilibration dynamics} 
\begin{figure}[!htb]
\centering
\includegraphics[scale=0.5]{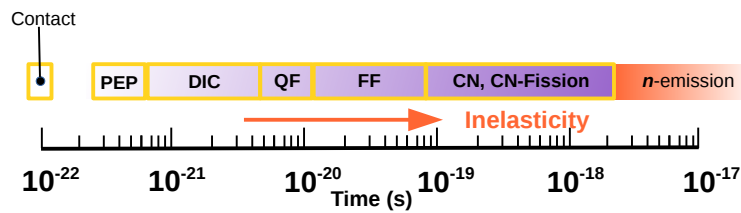}
\caption{Time-scales for various reaction types as a function of increasing inelasticity.\label{fig1}}
\end{figure}

In this section we discuss equilibration times for mass, isospin, and TKE in low-energy
heavy-ion reactions. Figure~\ref{fig1} shows the general time-scales associated with various
reaction types as a function of increasing inelasticity. It is important to state that somewhere
in the quasifission time-scale range (slow quasifission)~\cite{khuyagbaatar2015,khuyagbaatar2018}
we observe a total TKE loss of the reaction products, TKE following the Viola systematics.
In this sense quasifission reactions are most suitable for the study of time-scales for mass
equilibration, whereas time-scales for isospin and TKE equilibration can be investigated in
deep-inelastic reactions.

\subsection{Quasifission}
\begin{figure}[!htb]
\centering
\includegraphics[scale=0.35]{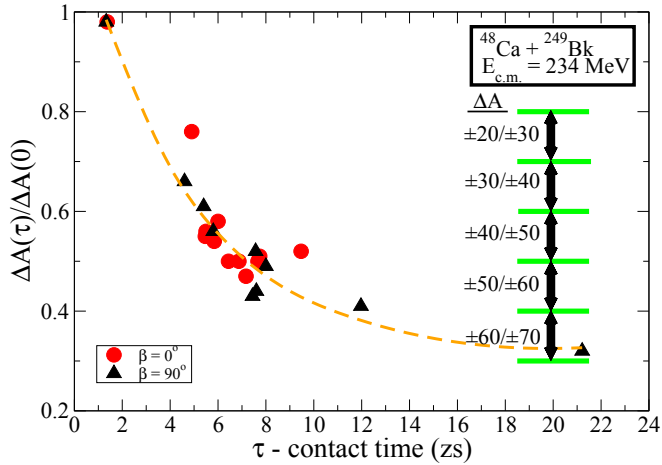}
\caption{The ratio of the final and initial fragment masses as a function of contact time for the $^{48}\mathrm{Ca}+^{249}\mathrm{Bk}$ system
at $E_{\mathrm{c.m.}}=234$~MeV and for two orientations of the $^{249}\mathrm{Bk}$ nucleus.
The dashed line shows one possible fit.\label{fig2}}
\end{figure}

\begin{figure}[!htb]
\centering
\includegraphics[scale=0.35]{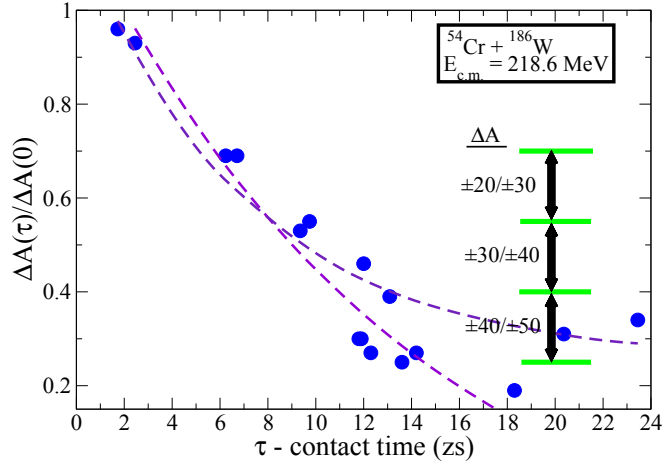}
\caption{The ratio of the final and initial fragment masses as a function of contact time for the $^{54}\mathrm{Cr}+^{186}\mathrm{W}$ system
at $E_{\mathrm{c.m.}}=218.6$~MeV and for two orientations of the $^{186}\mathrm{W}$ nucleus. Dashed lines show two possible fits.\label{fig3}}
\end{figure}

Quasifission, which occurs typically for systems with product of the charges $Z_1Z_2>1600$,
is characterized by two final
state fragments that emerge after a long lived composite system (typically longer
than 5\,zs) and final fragment masses $A_{f} = A_{\rm CN}/2 \pm 20$ or more,
and thus occupy the regime between quasielastic
and fusion/fission.
In addition, final TKE's distinguish quasifission from highly damped deep-inelastic
collisions, which have a smaller mass and charge difference between initial and
final fragments. In TDHF the mass and charge difference between the initial nuclei
and the final fragments measure the number of nucleons transferred.
In recent years a compelling number of TDHF calculations of quasifission have shown that TDHF is
an excellent predictor for the experimentally measured quantities, such as the
mass-angle distributions~\cite{golabek2009,kedziora2010,wakhle2014,oberacker2014,umar2015a,hammerton2015,umar2016,sekizawa2016,yu2017,morjean2017,wakhle2018}.
Due to the long contact times the quasifission process is suitable to study mass equilibration.
In Fig.~\ref{fig2} we plot the ratio of final and initial mass differences defined generally by,
\begin{eqnletter}
\Delta A(t) = A_{TLF}(t)-A_{PLF}(t)\;,
\label{dA}
\end{eqnletter}
as a function of contact time for the
$^{48}\mathrm{Ca}+^{249}\mathrm{Bk}$ system at $E_{\mathrm{c.m.}}=234$~MeV, and
for two extreme orientations of the deformed $^{249}\mathrm{Bk}$ nucleus indicated by the angle
$\beta$ which is the angle between the symmetry axis of the nucleus and the collision
axis. The points correspond the the impact parameters used, ranging from head-on collisions to more peripheral
collisions.
The horizontal bars on the right side of the figure indicate the number of particles
transferred between the target and the projectile. We observe that more mass transfer happens
at larger contact times as expected. The dashed line shows a typical fit of a function in the
form of $c_0+c_1exp(-\tau/\tau_0)$. Depending on the quality of the fit we obtain equilibration
times between $5-10$~zs.
In Fig.~\ref{fig3} we plot the same fragment mass difference ratio for the $^{54}\mathrm{Cr}+^{186}\mathrm{W}$ system at
$E_{\mathrm{c.m.}}=218.6$~MeV and for two orientations of the $^{186}\mathrm{W}$ nucleus.
On the plot we show two possible fits to the points. The obtained equilibration times range between $7-12$~zs.
More calculations are underway to study the orientation angle dependence of the equilibration times in more detail.
However, based on these and other results obtained from TDHF calculations we can safely conclude that mass equilibration
times are long, and a typical time of around 10~zs could be argued. Mass equilibration times can also be influenced by
shell effects. While the preference of Pb isotopes as quasifission product was theoretically observed in
TDHF calculations of $^{48}\mathrm{Ca}+^{238}\mathrm{U}$
system~\cite{hinde2012,durietz2013,wakhle2014,williams2018,mohanto2018}, it was recently confirmed experimentally by explicit charge measurement of the quasifission products~\cite{morjean2017} in the $^{48}\mathrm{Ti}+^{238}\mathrm{U}$ system. Naturally, influence of shell effects will depend
on the target projectile combinations but it is clear that shell effects do introduce a delay in the equilibration process.
\begin{figure}[!htb]
\centering
\includegraphics[scale=0.35]{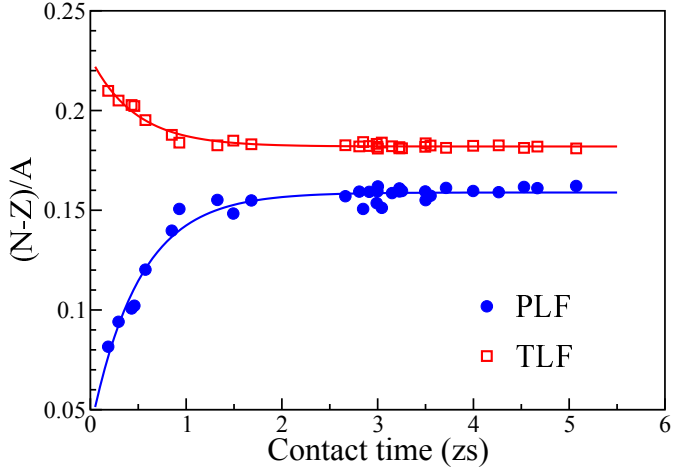}
\caption{The $(N-Z)/A$ value of the primary PLF (full circles) and TLF (open squares)
formed in $^{78}\mathrm{Kr}+^{208}\mathrm{Pb}$
at $E=8.5$~MeV/nucleon are plotted as a function of the contact time
between the collision partners.
The solid lines show fits to the TDHF results.\label{fig4}}
\end{figure}

\subsection{Deep-inelastic reactions}
Study of strongly damped collisions of nuclei or so called deep-inelastic collisions
can play an important role in elucidating the dynamics of charge and mass exchange,
dissipation of energy and angular momentum, degree of isospin equilibration,
and the dependence of these quantities on the properties of the reactants
such as the neutron-to-proton ratio ($N/Z$)\,\cite{schroder1977,moretto1981,toke1992}.
Here, we focus on the study of equilibration times for isospin and TKE.
Recently, an experimental study~\cite{jedele2017} in the
Fermi energy range have obtained an isospin equilibration time of about 0.3~zs. We have performed studies of the ${}^{78}\mathrm{Kr}+{}^{208}\mathrm{Pb}$
system at at 8.5~MeV/nucleon~\cite{umar2017}.
In Fig.~\ref{fig4} we plot the $(N-Z)/A$ value of the primary PLF (full circles) and TLF (open squares)
formed in $^{78}\mathrm{Kr}+^{208}\mathrm{Pb}$
at $E=8.5$~MeV/nucleon are plotted as a function of the contact time
between the collision partners. The solid lines show fits to the TDHF results.
For this system the mean life time of the
charge equilibration process, obtained from the final $(N-Z)/A$ value of the fragments is $\sim0.5$~zs. This and other studies suggest that for low-energy
heavy-ion collisions isospin equilibration occurs in the time-scale range of $0.5-1.0$~zs.
\begin{figure}[!htb]
    \centering
    \includegraphics[scale=0.35]{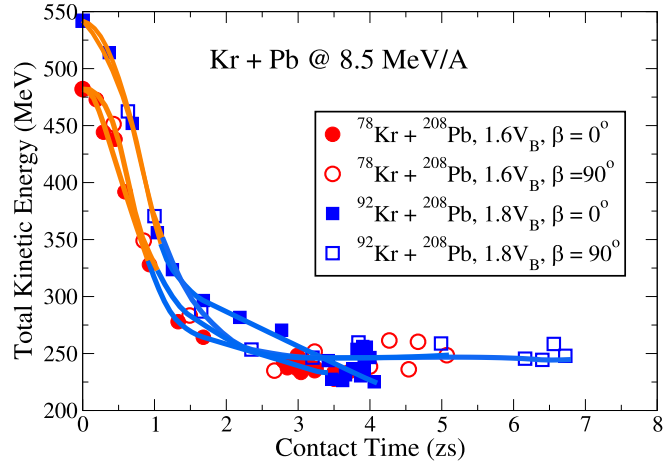}
    \caption{The exit channel TKE in $^{78}\mathrm{Kr}+^{208}\mathrm{Pb}$ reaction
        at $E=8.5$~MeV/nucleon are plotted as a function of the contact time
        between the collision partners.\label{fig5}}
\end{figure}

In order to study TKE equilibration times in Fig.~\ref{fig5} we plot the exit channel TKE for the $^{78}\mathrm{Kr}+^{208}\mathrm{Pb}$ and $^{92}\mathrm{Kr}+^{208}\mathrm{Pb}$ reactions at $E=8.5$~MeV/nucleon as a function of the contact time.
The calculations are done for two extreme orientations of the deformed Kr nuclei. The points correspond to
calculations for impact parameters in the range of $0-10$~fm. We observe that for both systems the TKE
rapidly falls initially as a function of contact time but slowly stabilizes around and after the contact times $1.5-2$~zs.
From this we may conclude that the equilibration time for TKE is in the range of $1.5-2$~zs. Similar results have
been recently found for the Ni$+$Ni system~\cite{williams2018}.

\section{Summary}
We have presented a discussion of equilibration times for mass, isospin, and TKE using the TDHF
approach. We find that mass equilibration times are much longer than those for isospin and TKE.
The fully microscopic TDHF theory has shown itself to be rich in
nuclear phenomena and continues to stimulate our understanding of nuclear dynamics.
The time-dependent mean-field studies seem to show that the dynamic evolution
builds up correlations that are not present in the static theory.
Although there is evidence that one-body dissipation can properly account for the transport phenomena
seen in these reactions further experiments are needed to test this conclusion.
We plan to supplement the studies mentioned in this manuscript with more TDHF calculations
and provide a more detailed analysis of equilibration times in low-energy heavy-ion reactions.


\acknowledgments
This work has been supported by the U.S. Department of Energy under grant No.
DE-SC0013847 with Vanderbilt University and by the
Australian Research Council Grants No. FT120100760 and DP160101254.

\bibliography{VU_bibtex_master}

\begin{thebibliography}{10}
\expandafter\ifx\csname url\endcsname\relax\def\url#1{\texttt{#1}}\fi
\expandafter\ifx\csname urlprefix\endcsname\relax\def\urlprefix{URL }\fi

\bibitem{balantekin2014}
\NAME{Balantekin A.~B., Carlson J., Dean D.~J., Fuller G.~M., Furnstahl R.~J.,
  Hjorth-Jensen M., Janssens R. V.~F., Li B.-A., Nazarewicz W., Nunes F.~M.,
  Ormand W.~E., Reddy S. \atque Sherrill B.~M.}, \IN{Mod. Phys. Lett.
  A}{29}{2014}{1430010}.

\bibitem{jedele2017}
\NAME{Jedele A., {McIntosh} A.~B., Hagel K., Huang M., Heilborn L., Kohley Z.,
  May L.~W., {McCleskey} E., Youngs M., Zarrella A. \atque Yennello S.~J.},
  \IN{Phys. Rev. Lett.}{118}{2017}{062501}.

\bibitem{tsang2004}
\NAME{Tsang M.~B., Liu T.~X., Shi L., Danielewicz P., Gelbke C.~K., Liu X.~D.,
  Lynch W.~G., Tan W.~P., Verde G., Wagner A., Xu H.~S., Friedman W.~A.,
  Beaulieu L., Davin B., {de Souza} R.~T., Larochelle Y., Lefort T., Yanez R.,
  Viola V.~E., Charity R.~J. \atque Sobotka L.~G.}, \IN{Phys. Rev.
  Lett.}{92}{2004}{062701}.

\bibitem{planeta1988}
\NAME{Planeta R., Zhou S.~H., Kwiatkowski K., Wilson W.~G., Viola V.~E., Breuer
  H., Benton D., Khazaie F., {McDonald} R.~J., Mignerey A.~C., Weston-Dawkes
  A., {de Souza} R.~T., Huizenga J.~R. \atque Schr\"oder W.~U.}, \IN{Phys. Rev.
  C}{38}{1988}{195}.

\bibitem{desouza1988b}
\NAME{{de Souza} R.~T., Schr\"oder W.~U., Huizenga J.~R., Planeta R.,
  Kwiatkowski K., Viola V.~E. \atque Breuer H.}, \IN{Phys. Rev.
  C}{37}{1988}{1783}.

\bibitem{planeta1990}
\NAME{P\l{}aneta R., Kwiatkowski K., Zhou S.~H., Viola V.~E., Breuer H.,
  {McMahan} M.~A., Kehoe W. \atque Mignerey A.~C.}, \IN{Phys. Rev.
  C}{41}{1990}{942}.

\bibitem{danielewicz2002}
\NAME{Danielewicz P., Lacey R. \atque Lynch W.~G.},
  \IN{Science}{298}{2002}{1592}.

\bibitem{rizzo2008a}
\NAME{Rizzo J., Chomaz P. \atque Colonna M.}, \IN{Nucl. Phys.
  A}{806}{2008}{40}.

\bibitem{colonna1998}
\NAME{Colonna M., {Di Toro} M., Guarnera A., Maccarone S., Zielinska-Pfab\'{e}
  M. \atque Wolter H.~H.}, \IN{Nucl. Phys. A}{642}{1998}{449}.

\bibitem{toro2010}
\NAME{{Di Toro} M., Baran V., Colonna M. \atque Greco V.}, \IN{J. Phys.
  G}{37}{2010}{083101}.

\bibitem{rizzo2011}
\NAME{Rizzo C., Baran V., Colonna M., Corsi A. \atque {Di Toro} M.}, \IN{Phys.
  Rev. C}{83}{2011}{014604}.

\bibitem{zhang2012}
\NAME{Zhang Y., Coupland D. D.~S., Danielewicz P., Li Z., Liu H., Lu F., Lynch
  W.~G. \atque Tsang M.~B.}, \IN{Phys. Rev. C}{85}{2012}{024602}.

\bibitem{colonna2013}
\NAME{Colonna M.}, \IN{Phys. Rev. Lett.}{110}{2013}{042701}.

\bibitem{simenel2012}
\NAME{Simenel C.}, \IN{Eur. Phys. J. A}{48}{2012}{152}.

\bibitem{simenel2018}
\NAME{Simenel C. \atque Umar A.~S.}, \IN{Prog. Part. Nucl.
  Phys.}{103}{2018}{19}.

\bibitem{khuyagbaatar2015}
\NAME{Khuyagbaatar J., Hinde D.~J., Carter I.~P., Dasgupta M., D\"ullmann
  {\relax Ch. E}., Evers M., Luong D.~H., {du Rietz} R., Wakhle A., Williams E.
  \atque Yakushev A.}, \IN{Phys. Rev. C}{91}{2015}{054608}.

\bibitem{khuyagbaatar2018}
\NAME{Khuyagbaatar J., David H.~M., Hinde D.~J., Carter I.~P., Cook K.~J.,
  Dasgupta M., D\"ullmann {\relax Ch. E}., Jeung D.~Y., Kindler B., Lommel B.,
  Luong D.~H., Prasad E., Rafferty D.~C., Sengupta C., Simenel C., Simpson
  E.~C., Smith J.~F., Vo-Phuoc K., Walshe J., Wakhle A., Williams E. \atque
  Yakushev A.}, \IN{Phys. Rev. C}{97}{2018}{064618}.

\bibitem{golabek2009}
\NAME{{C\'edric Golabek} \atque {C\'edric Simenel}}, \IN{Phys. Rev.
  Lett.}{103}{2009}{042701}.

\bibitem{kedziora2010}
\NAME{{David J. Kedziora} \atque {C\'edric Simenel}}, \IN{Phys. Rev.
  C}{81}{2010}{044613}.

\bibitem{wakhle2014}
\NAME{Wakhle A., Simenel C., Hinde D.~J., Dasgupta M., Evers M., Luong D.~H.,
  du~Rietz R. \atque Williams E.}, \IN{Phys. Rev. Lett.}{113}{2014}{182502}.

\bibitem{oberacker2014}
\NAME{Oberacker V.~E., Umar A.~S. \atque Simenel C.}, \IN{Phys. Rev.
  C}{90}{2014}{054605}.

\bibitem{umar2015a}
\NAME{Umar A.~S., Oberacker V.~E. \atque Simenel C.}, \IN{Phys. Rev.
  C}{92}{2015}{024621}.

\bibitem{hammerton2015}
\NAME{Hammerton K., Kohley Z., Hinde D.~J., Dasgupta M., Wakhle A., Williams
  E., Oberacker V.~E., Umar A.~S., Carter I.~P., Cook K.~J., Greene J., Jeung
  D.~Y., Luong D.~H., {McNeil} S.~D., Palshetkar C.~S., Rafferty D.~C., Simenel
  C. \atque Stiefel K.}, \IN{Phys. Rev. C}{91}{2015}{041602(R)}.

\bibitem{umar2016}
\NAME{Umar A.~S., Oberacker V.~E. \atque Simenel C.}, \IN{Phys. Rev.
  C}{94}{2016}{024605}.

\bibitem{sekizawa2016}
\NAME{Sekizawa K. \atque Yabana K.}, \IN{Phys. Rev. C}{93}{2016}{054616}.

\bibitem{yu2017}
\NAME{{Chong Yu} \atque {Lu Guo}}, \IN{Sci. China Phys.}{60}{2017}{092011}.

\bibitem{morjean2017}
\NAME{Morjean M., Hinde D.~J., Simenel C., Jeung D.~Y., Airiau M., Cook K.~J.,
  Dasgupta M., Drouart A., Jacquet D., Kalkal S., Palshetkar C.~S., Prasad E.,
  Rafferty D., Simpson E.~C., Tassan-Got L., Vo-Phuoc K. \atque Williams E.},
  \IN{Phys. Rev. Lett.}{119}{2017}{222502}.

\bibitem{wakhle2018}
\NAME{Wakhle A., Hammerton K., Kohley Z., Morrissey D.~J., Stiefel K., Yurkon
  J., Walshe J., Cook K.~J., Dasgupta M., Hinde D.~J., Jeung D.~J., Prasad E.,
  Rafferty D.~C., Simenel C., Simpson E.~C., Vo-Phuoc K., King J., Loveland W.
  \atque Yanez R.}, \IN{Phys. Rev. C}{97}{2018}{021602}.

\bibitem{hinde2012}
\NAME{Hinde D.~J., du~Rietz R., Simenel C., Dasgupta M., Wakhle A., Evers M.
  \atque Luong D.~H.}, \IN{AIP Conf. Proc.}{1423}{2012}{65}.

\bibitem{durietz2013}
\NAME{du~Rietz R., Williams E., Hinde D.~J., Dasgupta M., Evers M., Lin C.~J.,
  Luong D.~H., Simenel C. \atque Wakhle A.}, \IN{Phys. Rev.
  C}{88}{2013}{054618}.

\bibitem{williams2018}
\NAME{Williams E., Sekizawa K., Hinde D.~J., Simenel C., Dasgupta M., Carter
  I.~P., Cook K.~J., Jeung D.~Y., McNeil S.~D., Palshetkar C.~S., Rafferty
  D.~C., Ramachandran K. \atque Wakhle A.}, \IN{Phys. Rev.
  Lett.}{120}{2018}{022501}.

\bibitem{mohanto2018}
\NAME{Mohanto G., Hinde D.~J., Banerjee K., Dasgupta M., Jeung D.~Y., Simenel
  C., Simpson E.~C., Wakhle A., Williams E., Carter I.~P., Cook K.~J., Luong
  D.~H., Palshetkar C.~S. \atque Rafferty D.~C.}, \IN{Phys. Rev.
  C}{97}{2018}{054603}.

\bibitem{schroder1977}
\NAME{Schr\"oder W.~U. \atque Huizenga J.~R.}, \IN{Annu. Rev. Nucl. Part.
  Sci.}{27}{1977}{465}.

\bibitem{moretto1981}
\NAME{Moretto L.~G. \atque Schmitt R.~P.}, \IN{Rep. Prog.
  Phys.}{44}{1981}{533}.

\bibitem{toke1992}
\NAME{T{\~{o}}ke J. \atque Schr\"oder W.~U.}, \IN{Annu. Rev. Nucl. Part.
  Sci.}{42}{1992}{401}.

\bibitem{umar2017}
\NAME{Umar A.~S., Simenel C. \atque Ye W.}, \IN{Phys. Rev.
  C}{96}{2017}{024625}.

\end{thebibliography}

\end{document}